\documentclass[twocolumn]{aastex61}

\received{July 1, 2016}
\revised{September 27, 2016}
\accepted{\today}
\submitjournal{ApJ}

\shorttitle{Ionization Response to LGRB's Prompt Emission}
\shortauthors{Wang et al.}
\begin{document}

\title{A Global Photoionization Response to Prompt Emission and Outliers: Different Origin of Long Gamma-ray Bursts?}

\correspondingauthor{J. Wang}
\email{wj@bao.ac.cn}

\author{J. Wang}
\affil{Key Laboratory of Space Astronomy and Technology, National Astronomical Observatories, Chinese Academy of Sciences, Beijing
100012, China}
\affil{School of Astronomy and Space Science, University of Chinese Academy of Sciences, Beijing, China}

\author{L. P. Xin}
\affil{Key Laboratory of Space Astronomy and Technology, National Astronomical Observatories, Chinese Academy of Sciences, Beijing
100012, China}

\author{Y. L. Qiu}
\affil{Key Laboratory of Space Astronomy and Technology, National Astronomical Observatories, Chinese Academy of Sciences, Beijing
100012, China}

\author{D. W. Xu}
\affil{Key Laboratory of Space Astronomy and Technology, National Astronomical Observatories, Chinese Academy of Sciences, Beijing
100012, China}
\affil{School of Astronomy and Space Science, University of Chinese Academy of Sciences, Beijing, China}

\author{J. Y. Wei}
\affiliation{Key Laboratory of Space Astronomy and Technology, National Astronomical Observatories, Chinese Academy of Sciences, Beijing
100012, China}
\affiliation{School of Astronomy and Space Science, University of Chinese Academy of Sciences, Beijing, China}



\begin{abstract}

By using the line ratio \ion{C}{4}$\lambda1549$/\ion{C}{2}$\lambda1335$ as a tracer of 
ionization ratio of the interstellar medium (ISM) illuminated by a long gamma-ray burst (LGRB), 
we identify a global photoionization response of the ionization ratio to the photon luminosity of the prompt emission 
assessed by either $L_{\mathrm{iso}}/E_{\mathrm{peak}}$ or $L_{\mathrm{iso}}/E^2_{\mathrm{peak}}$.  
The ionization ratio increases with both $L_{\mathrm{iso}}/E_{\mathrm{peak}}$ and 
$L_{\mathrm{iso}}/E^2_{\mathrm{peak}}$ for a majority of the LGRBs in our sample, although there are a few outliers.
The identified dependence of \ion{C}{4}/\ion{C}{2} on $L_{\mathrm{iso}}/E^2_{\mathrm{peak}}$ suggests that 
the scatter of the widely accepted Amati relation is related with the ionization ratio in ISM. 
The outliers tend to have 
relatively high \ion{C}{4}/\ion{C}{2} values as well as relatively high 
\ion{C}{4}$\lambda1549$/\ion{Si}{4}$\lambda1403$ ratios, which suggests an existence of
Wolf-Rayet stars in the environment of these LGRBs. We finally argue that the outliers and the LGRBs following the
identified  \ion{C}{4}/\ion{C}{2}$-L_{\mathrm{iso}}/E_{\mathrm{peak}}$ ($L_{\mathrm{iso}}/E^2_{\mathrm{peak}}$) 
correlation might come from different progenitors with different local environments.

\end{abstract}




\section{Introduction} \label{sec:intro}

Long gamma-ray bursts (LGRBs) are the most powerful explosions
occurring from local universe (the nearest one is GRB\,980425 at $z=0.008$, Galama et al. 1998) 
to very high redshift (e.g., Salvaterra et al. 2009; Tanvir et al. 2009).
Up to date, the most distant one reported in literature is GRB\,090429B with a photometric redshift of 
$z=9.4$ (Cucchiara et al. 2011). 
The detection of the associated supernova in a few LGRBs (see Cano et al. 2016 for a recent review) 
strongly supports that LGRBs originate from the core-collapse of young massive
stars ($\geq25M_\odot$) (e.g., Hjorth \& Bloom 2012; Woosley \& Bloom 2006
and references therein).
The GRB's afterglow at a wide wavelength range from radio to X-ray is produced 
through the synchrotron radiation when the  
jet ignited in the core-collapse impacts and shocks the surrounding medium (e.g., 
Meszaros \& Rees 1997; Sari et al. 1998).

Within the first hours after the onset of a burst, the powerful afterglows of 
LGRBs illuminate not only the interstellar medium (ISM) of their host galaxies, but also the 
intergalactic medium, which produces an afterglow optical spectrum associated with multiple strong absorption lines of metals with
different ionization stages at different redshifts (e.g., Fybno et al. 2009; de Ugarte Postigo et al. 2012). The spectra provide
us an opportunity to study the properties 
of both medium at GRB's local environment and intervening absorption clouds located between
the host galaxy and the observer (e.g., Savaglio
et al. 2003; Butler et al. 2003; Prochaska et al. 2006, 2007; Tejos et al. 2007; Vergani et al. 2009;
Vreeswijk et al. 2007; Kawai et al. 2006; Totani et al. 2006; 
D'Elia et al. 2009, 2010; Wang 2013).  

A response of the host galaxy environment to both prompt and afterglow emission has been proposed and observed for a long time. 
The theoretical study in Perna et al. (2003) suggested that the silicates can be destroyed by 
the strong X-ray/UV radiation (see also in Waxman \& Draine 2000).  
The photoionization effect due to the prompt and afterglow emission of GRBs has been 
identified in the X-ray spectra of a few bursts (e.g., Amati et al. 2000, Antonelli et al., 2000; Piro et al. 2000). 
Due to the afterglow evolution, an evolution of 
photoionization of the medium around the progenitors has been put forward in Perna \& Loeb (1998), which is subsequently 
supported by the observed time variability of absorption lines in a few bursts (e.g., 
GRB\,010222, Mirabal et al. 2002; GRB\,020813, Dessauges-Zavadsky et al. 2006; GRB\,060418, Vreeswijk et al. 2007; 
GRB\,060206, Hao et al. 2007; GRB\,080310, Vreeswijk et al. 2013; GRB\,100901, Hartoog et al. 2013). 
In addition, with a sample of 69 low-resolution afterglow spectra, de Ugarte Postigo et al. (2012) 
revealed a global weak dependence of the ionization ratio quantified by the line strength parameter on the rest-frame  
isotropic prompt energy $E_{\mathrm{iso}}$ in 1-1000keV band.  

In this paper, we identify a global photoionization response of the ISM to GRB's prompt emission by 
instead focusing on the role of high energy ionizing photon flux. The result further motivates us to suspect that 
there are two kinds of origin of LGRBs. The paper is organized as follows. The sample selection is presented 
Section 2. Section 3 shows the results and implications.

\section{Sample} \label{sec:sample}

Our aim is to study the global photoionization response of the host galaxy ISM to LGRB's prompt emission.
We adopt the line ratio of \ion{C}{4}$\lambda\lambda$1548,1550 and \ion{C}{2}/\ion{C}{2}$^*\lambda\lambda$1334,1335 as a
tracer of ionization ratio of the ISM of individual LGRB, both because the two lines are usually quite strong in 
the afterglow spectra and because the ionization potential of \ion{C}{4} is as high as 47.89eV.
We compile a sample of \it Swift \rm (Gehrels et al. 2004) LGRBs with reported measurements of both two absorption lines and 
prompt emission. 
Our sample is mainly complied from de Ugarte Postigo et al. (2012) which published a sample of low-resolution 
afterglow spectra of 69 LGRBs. For the bursts with measurements (including both detection and upper limit) of 
both \ion{C}{4} and \ion{C}{2}, the common objects with a measurement of prompt emission is extracted from
Ghirlanda et al. (2017) and Nava et al. (2012).
The sample is finally composed of 20 LGRBs and is tabulated in Table 1, along with the
references. Columns (1) and (2) list the identification
and the measured redshift of each LGRB, respectively. The measured equivalent widths (EWs) of 
\ion{C}{2}$\lambda1335$, \ion{C}{4}$\lambda1549$ and \ion{Si}{4}$\lambda1403$ absorption lines are given in 
Columns (3), (4) and (5), respectively. The line ratio in logarithmic of \ion{C}{4}/\ion{C}{2} is listed
column (6). Columns (7) and (8) are the rest-frame isotropic prompt luminosity and peak energy based on the standard Band spectrum.   
All the errors reported in the table correspond to the
1$\sigma$ significance level after taking into account the proper error
propagation.

\begin{table*}[h!]
\centering
\scriptsize
\caption{Sample of \it Swift \rm LGRBs with Measurements of Both Absorption Lines and Prompt Emission.} \label{tab:decimal}
\begin{tabular}{ccccccccc}
\tablewidth{0pt}
\hline
\hline
GRB & $z_{\mathrm{GRB}}$ & EW(CIV$\lambda1549$)  &  EW(CII/CII$^*\lambda1335)$ & EW(SiIV$\lambda1549$)   & $\log(\mathrm{CIV/CII})$ & $\log L_{\mathrm{iso}}$ 
& $\log E_{\mathrm{peak}}$  &  References\\
    &                    &       \AA              &     \AA                    & \AA                     &          &$\mathrm{erg\ s^{-1}}$   & keV  & \\
(1) & (2) & (3) & (4) & (5) & (6) & (7) & (8) & (9) \\ 
\hline
050401  & 2.89  & $3.08\pm0.77$ & $2.57\pm0.26$ & $<1.53$  & $-0.08\pm0.12$  & $53.30\pm0.02$ &  $2.69\pm0.09$ & 1,5\\      
050908\tablenotemark{*}   & 3.34  & $0.14\pm0.04$ & $3.23\pm0.05$ & $0.59\pm0.04$ & $1.36\pm0.12$   & $51.92\pm0.07$ &  $2.29\pm0.08$ & 1,4\\
050922C & 2.22  & $0.31\pm0.06$ & $1.27\pm0.03$ & $0.43\pm0.33$ & $0.61\pm0.08$   & $53.28\pm0.01$ &  $2.62\pm0.12$ & 1,4\\
060124  & 2.3   & $<0.7$        & $2.93\pm0.22$ & $<0.5$ & $>0.62$         & $53.15\pm0.01$ &  $2.80\pm0.11$ & 1,4\\
060210  & 3.91  & $4.32\pm0.08$ & $8.63\pm0.05$ & $3.05\pm0.06$ & $0.30\pm0.01$   & $52.78\pm0.06$ &  $2.76\pm0.14$ & 1,4\\
060714  & 2.71  & $2.83\pm0.13$ & $3.53\pm0.11$ & $1.56\pm0.11$ & $0.10\pm0.02$   & $52.15\pm0.03$ &  $2.37\pm0.20$ & 1,4\\
070110  & 2.35  & $1.13\pm0.09$ & $0.93\pm0.09$ & $0.42\pm0.09$  & $-0.08\pm0.05$  & $51.65\pm0.07$ &  $2.57\pm0.20$ & 1,4\\
070411\tablenotemark{*}  & 2.95  & $0.71\pm0.13$ & $2.56\pm0.13$ & $0.66\pm0.11$ & $0.56\pm0.08$   & $51.72\pm0.04$ &  $2.68\pm0.09$ & 1,4\\
071031  & 2.69  & $1.21\pm0.04$ & $2.18\pm0.04$ & $0.75\pm0.04$ & $0.26\pm0.02$   & $51.28\pm0.04$ &  $1.64\pm0.09$ & 1,4\\
080319C & 1.95  & $<2.88$       & $3.59\pm0.24$ & $<1.87$ & $>0.10$         & $52.98\pm0.01$ &  $3.24\pm0.13$ & 1,4\\
080603B & 2.69  & $1.05\pm0.04$ & $1.19\pm0.04$ & $0.43\pm0.04$ & $0.05\pm0.02$   & $53.08\pm0.02$ &  $2.58\pm0.09$ & 1,4\\
080605  & 1.64  & $<6.89$       & $5.11\pm0.30$ & \dotfill & $>-0.13$        & $53.51\pm0.02$ &  $2.82\pm0.03$ & 1,4\\
080721  & 2.59  & $1.98\pm0.19$ & $4.46\pm0.17$ & $0.84\pm0.04$ & $0.35\pm0.04$   & $54.01\pm0.06$ &  $3.24\pm0.06$ & 1,4\\
080810\tablenotemark{*}  & 3.35  & $<0.43$       & $3.13\pm0.21$ & $<0.21$ & $>0.86$         & $52.97\pm0.04$ &  $3.17\pm0.10$ & 1,4\\  
081008  & 1.97  & $1.41\pm0.10$ & $1.04\pm0.07$ & $0.44\pm0.07$ & $-0.13\pm0.04$  & $51.48\pm0.04$ &  $2.43\pm0.09$ & 1,4\\
080928  & 1.69  & $<1.44$       & $2.12\pm0.26$ & $<0.26$ & $>0.17$         & $51.48\pm0.04$ &  $2.30\pm0.09$ & 1,4\\
090516  & 4.11  & $5.81\pm0.04$ & $5.87\pm0.10$ & $1.72\pm0.04$ & $0.00\pm0.01$   & $52.86\pm0.02$ &  $2.97\pm0.34$ & 1,4\\
090812  & 2.45  & $1.94\pm0.21$ & $1.12\pm0.08$ & $0.57\pm0.12$ & $-0.24\pm0.06$  & $52.98\pm0.04$ &  $3.31\pm0.14$ & 1,4\\
110205  & 2.22  & $1.31\pm0.25$ & $0.73\pm0.20$ & \dotfill & $-0.25\pm0.14$  & $52.40\pm0.06$ &  $2.85\pm0.15$ & 2,4\\ 
140624  & 2.28  & $1.80\pm0.31$ & $5.02\pm0.21$ & $2.01\pm0.25$ & $0.45\pm0.08$   & $53.30\pm0.02$ &  $2.45\pm0.09$ & 3\\
\decimals
\hline
\end{tabular}
\tablecomments{References in the last column: (1) de Ugarte Postigo et al. 2012; (2) Cucchiara et al. 2011; (3) Xin et al. 2017;
(4) Ghirlanda et al. 2017; (5) Nava et al. 2012.  The three outliers in Figure 1 are marked with asterics in Column (1). 
}
\end{table*}

\section{Results and Discussion}

\subsection{\ion{C}{4}/\ion{C}{2}: A Global Photoionization Response of ISM to LGRBs' Prompt Emission} \label{subsec:tables}

A global ionization response of ISM to the prompt emission of LGRBs is shown in Figure 1 in which 
the line ratio of \ion{C}{4}/\ion{C}{2} is used as a tracer of ionization ratio of the ISM within the 
line-of-sight of an observer.  In stead of using $E_{\mathrm{iso}}$ as
an indicator of the strength of the prompt emission in the study of de Ugarte Postigo et al. (2012),   
the left panel in Figure 1 plots \ion{C}{4}/\ion{C}{2} as a function of $L_{\mathrm{iso}}/E_{\mathrm{peak}}$. 
Based on the definition of the Band spectrum, $L_{\mathrm{iso}}/E_{\mathrm{peak}}$ is 
the isotropic photon numbers emitted per second at the characteristic photon energy defined as $E_{\mathrm{peak}}$,
which is equivalent to the ionization parameter that is widely used in the photoionization models (e.g., Osterbrock \& Ferland 2006) if 
the densities of the ISM of different LGRBs are comparable.

For the bursts with a detection of EW of \ion{C}{2}, one can see that there is a dependence of \ion{C}{4}/\ion{C}{2} ratio on 
$L_{\mathrm{iso}}/E_{\mathrm{peak}}$, except for two outliers with relatively large \ion{C}{4}/\ion{C}{2} ratios.
Generally speaking, higher the \ion{C}{4}/\ion{C}{2} ratio, larger the $L_{\mathrm{iso}}/E_{\mathrm{peak}}$ will be, which 
indicates that the ionization ratio of the ISM around the bursts increases with the ionizing photon flux assessed from the 
prompt emission of LGRBs. 
A statistical test yields a Kendall's $\tau=0.238$ and a Z-value of 1.237 at a significance
level with a probability of null correlation of $P=0.216$. The significance of the dependence is 
considerably enhanced to be $\tau=0.539$, $Z=2.562$ and $P=0.0104$ when the two outliers are 
excluded from the statistics.
The significance is further enhanced obviously in the right panel of Figure 2, which plots 
line ratio \ion{C}{4}/\ion{C}{2} as a function of $L_{\mathrm{iso}}/E^2_{\mathrm{peak}}$.  
The physical meaning of $L_{\mathrm{iso}}/E^2_{\mathrm{peak}}$ can be understood as the specific photon  
numbers emitted per second with a photon energy of $E_{\mathrm{peak}}$. 
The same statistical test results in a significantly improved statistics with a $\tau=0.352$, $Z=1.836$ and $P=0.0664$ and   
a $\tau=0.603$, $Z=2.873$ and $P=0.0041$ when the outliers are excluded. 

Both \ion{C}{4}/\ion{C}{2}$-L_{\mathrm{iso}}/E_{\mathrm{peak}}$ and \ion{C}{4}/\ion{C}{2}$-L_{\mathrm{iso}}/E^2_{\mathrm{peak}}$
correlations suggest a photoionization effect in which the circumburst medium in the line-of-sight is photoionized by the   
GRB's prompt and afterglow emission. In fact, the photoionization effect is revealed in some previous case studies focusing on individual
GRBs. A transient absorption edge at $\sim3.8$keV, which is produced by the 
the circumburst medium highly ionized by the  GRB's prompt emission, is discovered in 
the X-ray spectrum of the prompt
emission of GRB\, 990705 (Amati et al. 2000). Emission features (e.g., Fe K$\alpha$ and Ly$\alpha$ lines) resulted from
photoionization by the GRB's prompt and afterglow emission is identified in the X-ray afterglow spectra of a few GRBs
(e.g., Antonelli et al., 2000; Piro et al. 1999, 2000, Yoshida et al. 1999).

The dependence of \ion{C}{4}/\ion{C}{2} on $L_{\mathrm{iso}}/E^2_{\mathrm{peak}}$ shown in the right panel in Figure 1 is quite interesting. 
In fact, previous 
statistical studies firmly established a tight correlation between $L_{\mathrm{iso}}$
and $E_{\mathrm{peak}}$, which results in a relationship of $L_{\mathrm{iso}}\propto E^2_{\mathrm{peak}}$ 
in both homogeneous and wind ISM
(e.g., Amati et al. 2002; Yonetoku et al. 2004; Ghirlanda et al. 2010, 2017; Nava et al. 2012). Subsequent 
studies suggested that the relationship is physically driven by the initial Lorentze factor (e.g., Nava et al. 2012; 
Ghirlanda et al. 2017 and references therein). The  dependence revealed by us therefore suggests that the 
scatter of the $L_{\mathrm{iso}}-E_{\mathrm{peak}}$ relationship is related with the \ion{C}{4}/\ion{C}{2} ratio.

A linear fitting FITEXY with uncertainties in both x and y coordinates yields a relationship of
\begin{equation}
 \log\frac{L_{\mathrm{iso}}}{E_{\mathrm{peak}}^2}=(47.15\pm0.07)+(2.38\pm0.28)\log\frac{\mathrm{CIV}}{\mathrm{CII}}
\end{equation}
The  D'Agostini fitting method (D'Agostini 2005, see also in e.g., Guidorzi et al. 2006, Amati et al. 2008) 
is alternatively used to model the linear relationship as $y=\beta_0+\beta_1x+\epsilon$,
where $\epsilon$ is the extra Gaussian scatter. The optimal values obtained by the 
maximum likelihood method (MLM) results in a relationship 
\begin{equation}
  \log\frac{\mathrm{CIV}}{\mathrm{CII}}=(-12.82\pm2.00)+(0.27\pm0.04)\log\frac{L_{\mathrm{iso}}}{E_{\mathrm{peak}}^2}
\end{equation}
with an extra scatter of $\epsilon=0.19$. Both best-fitted relationships are overplotted in the right panel of Figure 1.

Figure 2 shows the $E_{\mathrm{peak}}$ versus $L_{\mathrm{iso}}$ ($E_{\mathrm{iso}}$) correlation for the 
sample used in this study. The same D'Agostini fitting method returns best fits: 
\begin{equation}
 \log E_{\mathrm{peak}}=(-14.91\pm1.64)+(0.33\pm0.03)\log L_{\mathrm{iso}}
\end{equation}
associated with an extra scatter of $\epsilon=0.32$, and 
\begin{equation}
 \log E_{\mathrm{peak}}=(-26.13\pm2.86)+(0.54\pm0.05)\log E_{\mathrm{iso}}
\end{equation}
associated with an extra scatter of $\epsilon=0.27$. A comparison of the obtained scatters enables one to definitely
see that the dispersion of the $L_{\mathrm{iso}}/E_{\mathrm{Peak}}^2$ versus photoionization ratio correlation  
is smaller than both of the $E_{\mathrm{peak}}-L_{\mathrm{iso}}$ and $E_{\mathrm{peak}}-E_{\mathrm{iso}}$
correlations.

\begin{figure}[ht!]
\plotone{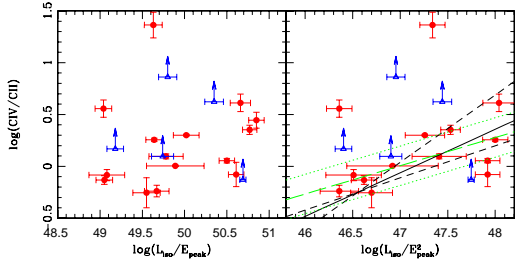}
\caption{\it Left panel: \rm line ratio \ion{C}{4}/\ion{C}{2} plotted against $L_{\mathrm{iso}}/E_{\mathrm{peak}}$.
The bursts with a determination of EW of \ion{C}{2} are shown by the red solid circles, and the bursts with an
upper limit of EW of \ion{C}{2}  by the blue open triangles and arrows. The errorbars correspond to the 1$\sigma$ significance level.
\it Right panel: \rm the same as the left one but for $L_{\mathrm{iso}}/E^2_{\mathrm{peak}}$. The black solid line shows the best fit for the 
\ion{C}{4}/\ion{C}{2} versus $L_{\mathrm{iso}}/E^2_{\mathrm{peak}}$ sequence through the FITEXY method (i.e., Eq.(1)). The 3$\sigma$ deviation from the best fit 
in both intercept and slope is shown by the black short-dashed lines. The green long-dashed and dotted lines presents the best fit and 1$\sigma$ deviations,
respectively, which is obtained through the D'Agostini fitting method (i.e., Eq.(2)).}
\end{figure}

\begin{figure}[ht!]
\plotone{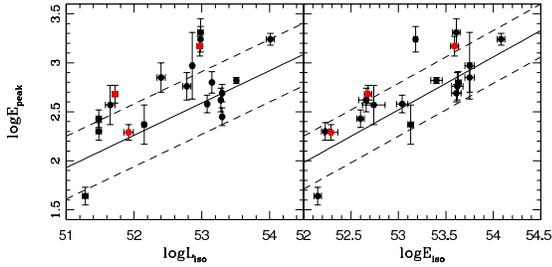}
\caption{\it Left panel: $E_{\mathrm{peak}}$ plotted against $L_{\mathrm{iso}}$. The best fitted linear relationship obtained through
the D'Agostini fitting method is presented by the solid line, and the 1$\sigma$ scatter by the two dashed lines. The three outliers 
(see Table 1 and Section 3.2) are marked by the red dots.
\it Right panel: the same as the left one but for $E_{\mathrm{iso}}$. }
\end{figure}

\subsection{Outliers: Different Origin of LGRBs?} 

By including both outliers and bursts with an upper limit of EW of \ion{C}{2}, 
the distribution on the \ion{C}{4}/\ion{C}{2} versus $L_{\mathrm{iso}}/E_{\mathrm{peak}}$ ($L_{\mathrm{iso}}/E^2_{\mathrm{peak}}$) diagram
further suggests 
that the bursts listed in our sample could be divided into two groups: a majority of the bursts that follow 
the ionization ratio versus ionizing photon flux dependence and a few outliers with relatively either large \ion{C}{4}/\ion{C}{2} ratios
or small $L_{\mathrm{iso}}/E_{\mathrm{peak}}$ ($L_{\mathrm{iso}}/E^2_{\mathrm{peak}}$) (or both). The three outliers in Figure 1 are: 
GRB\,050908, GRB\,070411 and GRB\,080810. 

In the first case, Martone et al. (2017) argued that the outliers in the $E_{\mathrm{peak}}-E_{\mathrm{iso}}$ correlation is possibly due to
an overestimation of $E_{\mathrm{peak}}$ resulted from an underestimation of X-ray prompt emission. In the current sample, Figure 2, however, 
shows that all the three outliers generally follow the fitted $E_{\mathrm{peak}}-E_{\mathrm{iso}}$ correlation, which suggests that 
the observational bias could be not a favorite explanation for the outliers.   

\rm
Alternatively, the large \ion{C}{4}/\ion{C}{2} ratios suggest a abnormally high ionization ratio in the ISM for the outliers.  A possible explanation of the high ionization ratio is an additional
photoionization contributed by the underlying intensive starformation, especially the massive Wolf-Rayet (WR) stars in the environment of 
the LGRBs. Taking into account the WR \rm outflow model, Berger et al. (2006) argued that the 
\ion{C}{4}$\lambda1549$/\ion{Si}{4}$\lambda1403$ line ratio is a good tracer of an existence of WRs because the outflow from the massive stars would increase 
the carbon metallicity in the ISM. In fact, a tendency of higher \ion{C}{4}/\ion{Si}{4} ratio in LGRBs than in QSOs is tentatively 
revealed in de Ugarte Postigo et al. (2012) through a large sample. 

In order to check if the WR star scenario is working for the outliers, Figure 2 shows the
cumulative distribution of \ion{C}{4}/\ion{Si}{4} ratio for the current sample.  The vertical lines mark the 
 \ion{C}{4}/\ion{Si}{4} values for the three outliers with the largest deviation from the \ion{C}{4}/\ion{C}{2} versus $L_{\mathrm{iso}}/E_{\mathrm{peak}}$
($L_{\mathrm{iso}}/E^2_{\mathrm{peak}}$) 
sequence. Clearly, both outliers with a firm measurement of EW of \ion{C}{2} have quite high \ion{C}{4}/\ion{Si}{4} ratios, which agrees with 
the prediction of the WR scenario quite well.  
  
\begin{figure}[ht!]
\plotone{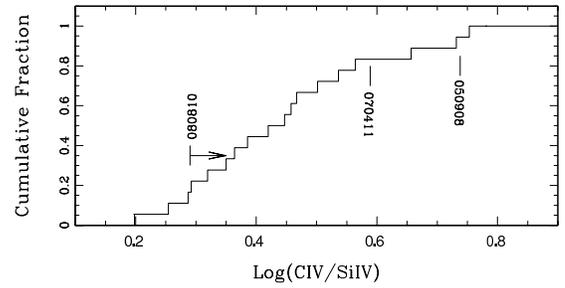}
\caption{Cumulative distribution of \ion{C}{4}/\ion{Si}{4} line ratio. The values of the three outliers in Figure 1 are marked by 
the vertical lines and labels.}
\end{figure}

The outliers with a property of both large \ion{C}{4}/\ion{C}{2} and \ion{C}{4}/\ion{Si}{4} ratios motivates us to suspect that the bursts following and deviating the 
\ion{C}{4}/\ion{C}{2}$-L_{\mathrm{iso}}/E_{\mathrm{peak}}$ ($L_{\mathrm{iso}}/E^2_{\mathrm{peak}}$) sequence are produced within different environments, 
or on other worlds, produced
by different progenitors. In fact, on the theoretical ground, both single-star 
model with a central engine of either a blackhole or a magentar (e.g., Woosley \& Heger 2006; Dai \& Lu 1998;  Zhang \&
Dai 2010; Wang et al. 2017)
and close interacting binary models (e.g.,  Fryer et al. 2007; van den Heuvel et al. 2007) have been proposed as 
the origin of LGRBs . 
On the observational ground, features emitted from WR stars have been detected in the spectra of the host galaxies 
of 8 nearby LGRBs (Han et al. 2010). With the measurements of metallicity of 
host galaxies of LGRBs up to $z\sim2$, a metallicity threshold of $Z_{\mathrm{th}}=0.7Z_\odot$ is suggested for the origin of LGRBs (e.g., 
Japelj et al. 2015; Vergani et al, 2017 and references therein). This threshold is, however, higher than the requirement of $0.2Z_\odot$ of the single-star model, which    
suggests an alternative progenitor of close interacting binary for some LGRBs, because the  binary model is less sensitive to 
the metallicity of the progenitor.

\section{Conclusion}

A correlation between line ratio \ion{C}{4}/\ion{C}{2} and $L_{\mathrm{iso}}/E_{\mathrm{peak}}$ ($L_{\mathrm{iso}}/E^2_{\mathrm{peak}}$)
is identified for a majority of LGRBs listed in this study, which suggests a 
global response of the ionization ratio to the ionizing photon luminosity assessed from the prompt emission.
The outliers of the correlation, which have both high \ion{C}{4}/\ion{C}{2} and \ion{C}{4}/\ion{Si}{4} ratios, motivate us 
to suspect that their progenitors differs from the bursts following the identified correlation.

%\clearpage

\acknowledgments

The authors would like to thank the anonymous referee for his/her careful review and helpful suggestions improving the 
manuscript.
JW \& DWX are supported by National Natural Science
Foundation of China under grants 11473036 and 11773036.
The study is supported by the National Basic Research Program of China
(grant 2014CB845800) and by the Strategic Pionner Program on Space Science, Chinese Academy of Sciences (Grant No.XDA15052600).

\end{document}